\input harvmac
\overfullrule=0pt
\def\half{{\textstyle{1\over 2}}}

\def\neqno#1{\eqnn{#1} \eqno #1}

\Title{hep-ph/0003173, UTPT-00-05} {\vbox{
\centerline{Brane Cosmologies without Orbifolds} }}
\centerline{\bf Hael Collins\footnote{$^\dagger$}{{\tt 
hael@physics.utoronto.ca}} and Bob 
Holdom\footnote{$^\ddagger$}{{\tt bob.holdom@utoronto.ca}} }
\medskip
\vbox{\it \centerline{Department of Physics}\vskip-2pt
\centerline{University of Toronto}\vskip-2pt
\centerline{Toronto, Ontario M5S 1A7, Canada}}

\bigskip
\bigskip
\centerline{Abstract}

\medskip {\baselineskip=10pt\centerline{\vbox{\hsize=.8\hsize\ninerm\noindent  
We study the dynamics of branes in configurations where 1) the brane is the
edge of a single anti-de Sitter (AdS) space and 2) the brane is the surface of
a vacuum bubble expanding into a Schwarzschild or AdS-Schwarzschild bulk.  In
both cases we find solutions that resemble the standard Robertson-Walker
cosmologies although in the latter, the evolution can be controlled by a mass
parameter in the bulk metric.  We also include a term in the brane action for
the scalar curvature.  This term adds a contribution to the low energy theory
of gravity which does not need to affect the cosmology but which is necessary
for the surface of the vacuum bubble to recover four dimensional gravity.  }}}

\Date{March, 2000}
\baselineskip=12pt

\newsec{Introduction.}

A remarkable feature in certain theories with more than the observed $3+1$
dimensions is that while these extra dimensions can extend infinitely, the
geometry of the bulk space-time nevertheless is able to confine gravity to a
three dimensional surface within the larger space.  Randall and Sundrum (RS)
\ref\rsb{L.~Randall and R.~Sundrum, ``An alternative to compactification,''
Phys.\ Rev.\ Lett.\ {\bf 83}, 4690 (1999) [hep-th/9906064].}
first showed that by attaching two semi-infinite slices of $4+1$ dimensional
anti-de Sitter space (AdS$_5$) along a three dimensional hypersurface, or
`$3$-brane', with orbifold conditions about this $3$-brane, gravity behaves as
though it is confined to its vicinity.  This $3$-brane is identified with our
universe.  In addition to reproducing ordinary Newtonian gravity, any
successful model should also be able to produce a realistic cosmological
evolution for the $3$-brane.  The dynamical evolution of the brane is
determined by Einstein's equations for the combined bulk and brane system, but
these equations might not produce the familiar Robertson-Walker cosmology
along the brane.  Viewed locally, near the brane the surrounding bulk
introduces a new element into the field equations for gravity on the brane
through a term for the change in the extrinsic curvature across the brane, as
originally derived by Israel
\ref\israelref{W.~Israel, ``Singular Hypersurfaces And Thin Shells In General
Relativity,'' Nuovo Cim.\ {\bf B44}, 1 (1966). }.
While generalizations of the original RS orbifold $\rsb$ have been shown to
admit the usual open, flat and closed Robertson-Walker cosmologies
\ref\per{P.~Kraus, ``Dynamics of anti-de Sitter domain walls,'' JHEP {\bf
9912}, 011 (1999) [hep-th/9910149]. },   
we shall examine more asymmetric geometries for which the AdS curvature
lengths on opposite sides of the brane are not necessarily equal.  We shall
treat in detail the case of a finite and spherical region of AdS space,
including the case of a vacuum bubble that expands in an asymptotically flat
$4+1$ dimensional space.

Most previous studies $\per$ 
\ref\garriga{J.~Garriga and M.~Sasaki, ``Brane-world creation and
black holes,'' hep-th/9912118.}
\ref\terningetal{C.~Csaki, M.~Graesser, C.~Kolda and J.~Terning, Phys.\ Lett.\ 
{\bf B462}, 34 (1999) [hep-ph/9906513]; 
C.~Csaki, M.~Graesser, L.~Randall and J.~Terning, hep-ph/9911406; 
J.~M.~Cline, C.~Grojean and G.~Servant, Phys.\ Rev.\ Lett.\ {\bf 83}, 4245
(1999) [hep-ph/9906523]; 
E.~E.~Flanagan, S.~H.~Tye and I.~Wasserman, hep-ph/9910498; 
P.~Binetruy, C.~Deffayet and D.~Langlois, hep-th/9905012; and 
T.~Shiromizu, K.~Maeda and M.~Sasaki, gr-qc/9910076.}
of the dynamics of a brane have only included a surface tension term and a
Lagrangian for the matter fields, which generally includes all the Standard
Model fields, in the brane action.  Yet without a more fundamental description 
of the physics that produces the brane, these terms should represent only the
leading pieces of an effective action
\ref\rseff{R.~Sundrum, ``Effective field theory for a three-brane universe,''
Phys.\ Rev.\ {\bf D59}, 085009 (1999) [hep-ph/9805471].}  
that could include higher order terms in a derivative expansion, such as a
term for the scalar curvature on the brane, ${\cal R}$, and higher powers of
curvature tensors on the brane, such as ${\cal R}^2$ and ${\cal R}_{ab}{\cal
R}^{ab}$.  Such terms generically are suppressed by extra powers of the AdS
curvature length scale, $\ell$, so at distances much larger than $\ell$ we
expect that these higher order terms in the brane action can be neglected. 
However, at least one fine-tuning is typically made to obtain a vanishing
cosmological constant on the brane by cancelling the brane tension against a
contribution from the bulk.  After this fine-tuning is made, a scalar
curvature term on the brane can be naturally of the same order as the terms
that remain in the field equations for gravity on the brane.  The importance
of such a term increases when we consider universes very different from the
original Randall-Sundrum scenario.  For a vacuum bubble in a asymptotically
flat bulk, this term is the sole source for four dimensional gravity.

A brane action that contains powers of the brane curvature tensors has also
been used in the context of the AdS/CFT correspondence
\ref\adscft{For a review, see O.~Aharony, S.~S.~Gubser, J.~Maldacena,
H.~Ooguri and Y.~Oz, ``Large N field theories, string theory and gravity,''
Phys.\ Rept.\  {\bf 323}, 183 (2000) [hep-th/9905111].}  
to regularize the action of a bulk AdS$_{n+1}$ space which diverges when the
radius of the AdS$_{n+1}$ space becomes infinite 
\ref\hs{M.~Henningson and K.~Skenderis, ``The holographic Weyl anomaly,'' JHEP
{\bf 9807}, 023 (1998) [hep-th/9806087].}  
\ref\bk{V.~Balasubramanian and P.~Kraus, ``A stress tensor for anti-de Sitter
gravity,'' Commun.\ Math.\ Phys.\  {\bf 208}, 413 (1999) [hep-th/9902121].}  
\ref\myers{R.~Emparan, C.~V.~Johnson and R.~C.~Myers, ``Surface terms as
counterterms in the AdS/CFT correspondence,'' Phys.\ Rev.\  {\bf D60}, 104001
(1999) [hep-th/9903238].}.  
Unlike the effective field theory description of the brane action, the
requirement that the total action of the theory---the sum of the brane action
and the bulk action---be finite in this limit precisely fixes the coefficients
of the terms in the brane action.  The coefficient of the brane tension gives
the usual cancellation of the cosmological constant on the brane; however, we
find that for the specific coefficient of the scalar curvature term on the
brane, the brane curvature term cancels the leading order effects coming from
the bulk gravity.  In the light of the AdS/CFT correspondence, this result
might be anticipated since the bulk gravitational theory is conjectured to be
equivalent to a conformal quantum field theory---without gravity---on the
surface.

In the next section, we derive the equations of motion for a bulk AdS$_{n+1}$
space in which a hypersurface is embedded.  In section three, these equations
are used to study a dynamic brane on which the induced metric takes the
standard Robertson-Walker form.  The general equations for a non-orbifolded
geometry including the effects of a scalar curvature term in the brane action
are found. Since these equations are difficult to solve exactly, in section
four we neglect the scalar curvature term and focus on the expansion of a
vacuum bubble in an asymptotically flat $4+1$ dimensional space-time.  In
section five we include the scalar curvature in the brane action and study its
effects when the brane is the edge of a single AdS space.  Finally we consider
its effect on the vacuum bubble, before concluding in section six.

\newsec{The Action for AdS$_{n+1}$ with a Boundary.}

We would like to derive the form of Einstein's equations on an $n$-dimensional
hypersurface embedded in an $n+1$-dimensional bulk space-time.  Later, we
restrict to the interesting case where $n=4$.  To be general, we shall treat
the bulk space-time as two regions, ${\cal M}_1$ and ${\cal M}_2$, separated
by the hypersurface, ${\cal B}$.  Note that these bulk regions do not need to
have the same metric on either side of the brane but only need to satisfy the
Israel conditions derived below.  Since the boundary corresponds to the
observed universe, we include an action on the brane containing, in addition
to a surface tension term, a term for the scalar curvature on the brane plus
the contributions from matter and gauge fields confined to the brane.  At each
point on the brane, we define a space-like unit normal, $N_a = N_a(x)$, to the
surface that satisfies $g^{ab}N_aN_b=1$.  $g_{ab}$ is the bulk metric and the
indices $a, b$ run over all the bulk coordinates.  The bulk metric induces a
metric on the brane,
$$h_{ab} = g_{ab} - N_a N_b; \neqno\defh$$ 
while the bulk metric can be discontinuous across the brane, the induced
metric on the brane should be the same whether calculated with the bulk metric
for either region.

Combining all of these ingredients, the total action is the sum of the actions
for the two bulk regions,\foot{Our convention for the sign of the Riemann
tensor is $-R^a_{\ bcd} \equiv \partial_d \Gamma^a_{bc} - \partial_c
\Gamma^a_{bd} + \Gamma^a_{ed}\Gamma^e_{bc} - \Gamma^a_{ec}\Gamma^e_{bd}$.}
$$\eqalign{ S_1 &= {1\over 16\pi G} \int_{{\cal M}_1} d^{n+1}x\, 
\sqrt{-g}\, \left[ R + {n(n-1)\over \ell_1^2} \right]  
- {1\over 8\pi G} \int_{\cal B} d^nx\, \sqrt{-h}\, K^{(1)} , \cr
S_2 &= {1\over 16\pi G} \int_{{\cal M}_2} d^{n+1}x\, \sqrt{-g}\, 
\left[ R + {n(n-1)\over \ell_2^2} \right] - {1\over 8\pi G} 
\int_{\cal B} d^nx\, \sqrt{-h}\, K^{(2)}  \cr}
\neqno\actionbulk$$ 
and that of the boundary,
$$S_{\rm surf} = {1\over 16\pi G} \int_{\cal B} d^nx\, \sqrt{-h}\, 
\left[ - {2(n-1)\over\ell} {\sigma\over \sigma_c} + b {\ell\over n-2} 
{\cal R} + 16\pi G\, {\cal L}_{\rm fields} + \cdots \right] . 
\neqno\actionsurf$$ 
Here $G$ is the bulk Newton's constant and $K$ is the trace of the extrinsic
curvature $K_{ab}$, defined by
$$K_{ab} = h_a^{\ c}\nabla_c N_b .\neqno\extrinsic$$ 
$\sigma$, ${\cal R}$ and ${\cal L}_{\rm fields}$ represent the brane tension,
the scalar curvature of the {\it induced\/} metric and the Lagrangian of
fields confined to the brane.  We normalize the brane tension with respect to
a critical tension,\foot{Note that since we are considering more general
geometries than the orbifolds of $\per$, it is convenient to define the
critical tension to be half that of the RS universe.} $\sigma_c = 3/8\pi
G\ell$, as will be useful later, and we allow the two bulk regions to have
potentially different curvature lengths $\ell_1$ or $\ell_2$.  This action is
a generalization of that which appears in $\bk$ and $\myers$.  From the
vantage of writing an effective theory on the brane $\rseff$, we simply
include the $\sqrt{-h}{\cal R}$ term as the next to leading term in the brane
action in powers of derivatives.  The coefficient of this term, $b\ell/(n-2)$,
is determined by some underlying theory so we leave it unspecified.

Varying the total action yields the usual Einstein equations in the bulk,
$$R_{ab} - \half R g_{ab} = {n(n-1)\over \ell_{1,2}^2} g_{ab} ,  
\neqno\eombulk$$ 
where the appropriate AdS length is chosen for each region, plus the following
equation for the surface,
$$\Delta K_{ab} - h_{ab} \Delta K = - {n-1\over\ell}{\sigma\over\sigma_c} 
h_{ab} - b {\ell\over n-2} \left[ {\cal R}_{ab} 
- {1\over 2}{\cal R} h_{ab} \right] + 8\pi G\, T_{ab} +
\cdots \neqno\jump$$ 
where $\Delta K_{ab} \equiv K^{(2)}_{ab} - K^{(1)}_{ab}$, $T_{ab}$ is the
energy-momentum tensor for the fields confined to the brane,
$$T_{ab} \equiv h_{ab}{\cal L}_{\rm fields} 
+ 2 {\delta{\cal L}_{\rm fields}\over\delta h^{ab}} , \neqno\defemt$$ 
and ${\cal R}_{ab}$ is the Ricci tensor for the induced metric.  Contracting
both sides of $\jump$ with $h^{ab}$ and solving for $\Delta K=h^{ab}\Delta
K_{ab}$ gives the Israel condition 
$$\Delta K_{ab} = {1\over\ell}{\sigma\over\sigma_c} h_{ab} 
- b {\ell\over n-2} \left[ {\cal R}_{ab} - {1\over 2(n-1)} 
{\cal R} h_{ab} \right] + 8\pi G \left[ T_{ab} 
- {1\over n-1} T^c_c h_{ab} \right] + \cdots \neqno\jumptwo$$ 
which describes the effect of the presence of the bulk space-time on the brane
Einstein equations through the appearance of the extrinsic curvature term.  A
similar equation, although without the term arising from varying the scalar
curvature in the brane action, has appeared in earlier studies of domain walls
in four
\ref\sikivie{J.~Ipser and P.~Sikivie, ``The Gravitationally Repulsive Domain
Wall,'' Phys.\ Rev.\  {\bf D30}, 712 (1984).}  
and five $\per$ $\garriga$ dimensions.  This term might seem unimportant at
distances much larger than $\ell$, since it contains two more powers of $\ell$
compared to the term with the brane tension.  However, the contributions from
$\sqrt{-h}{\cal R}$ can be of the same order as the difference between the
brane tension and $\Delta K_{ab}$, once the brane tension has been finely
tuned.

For comparison, in the original RS orbifold we only include the first term on
the right side of $\jumptwo$ and the extrinsic curvatures from the two sides
are equal and opposite, $K_{ab}\equiv K^{(1)}_{ab} = - K^{(2)}_{ab}$.  The
bulk AdS$_5$ space gives $K_{ab}=-h_{ab}/\ell$ which yields the usual
fine-tuning condition $\rsb$ for the brane tension, $\sigma = 2\sigma_c$.

\newsec{Cosmology on the Boundary.}

We shall now set $n=4$ and examine some specific solutions of the field
equations for gravity on an $3$-brane between two $4+1$ dimensional regions
with negative cosmological constants.  The metrics for the interior
$r<R(\tau)$ and exterior $r>R(\tau)$ regions with respect to the brane can be
written in the AdS$_5$-Schwarzschild form
\ref\db{D.~Birmingham, ``Topological black holes in anti-de Sitter space,''
Class.\ Quant.\ Grav.\  {\bf 16}, 1197 (1999) [hep-th/9808032].}.   
$$\eqalign{ ds^2\bigr|_{\rm int} &= - u(r)\, dt^2 + (u(r))^{-1}\, dr^2 
+ r^2\, d\Omega^2_3 \cr
u(r) &= {r^2\over\ell_1^2} + k - {m_1\over r^2} \cr 
ds^2\bigr|_{\rm ext} &= - v(r)\, dt^2 + (v(r))^{-1}\, dr^2 
+ r^2\, d\Omega^2_3 \cr 
v(r) &= {r^2\over\ell_2^2} + k - {m_2\over r^2} . \cr}
\neqno\adsbulk$$ 
An AdS$_5$ bulk corresponds to setting $m_1=m_2=0$.  We have included the
$-m_{1,2}/r^2$ terms in the metric since they can have an important effect on
the brane cosmology.  Their presence leads to black-hole horizons at some
distance into the bulk whose masses are determined by $m_1$ and $m_2$ $\db$ 
\ref\hawking{S.~W.~Hawking and D.~N.~Page, ``Thermodynamics Of Black Holes In
Anti-De Sitter Space,'' Commun.\ Math.\ Phys.\  {\bf 87}, 577 (1983). }.   

We shall frequently refer to the $k=1$ case, for which the brane is a
$3$-sphere and a closed Robertson-Walker cosmology results, but we shall leave
$k$ in the expressions with the understanding that flat or open cosmologies
can be obtained by setting $k=0$ or $-1$ respectively:  
$$d\Omega^2_3 \equiv \cases{ 
d\chi^2 + \sin^2\chi (d\theta^2 + \sin^2\theta\, d\phi^2) &$k=1$\cr 
\ell^{-2} (dx^2 + dy^2 + dz^2)  &$k=0$\cr  
d\chi^2 + \sinh^2\chi (d\theta^2 + \sin^2\theta\, d\phi^2) &$k=-1$\cr} 
\neqno\kcases$$ 

We are looking for dynamical solutions, so we let the position of the brane be
given by  
$$\left( t, r, \chi, \theta, \phi \right) 
= \left( T(\tau), R(\tau), \chi, \theta, \phi \right) \neqno\branelocation$$
where $\tau$ is the proper time for an observer at rest with respect to the
brane.  The normal to the brane is then 
$$N_a = \left( -\dot R, \dot T, 0, 0, 0 \right) \neqno\anormal$$ 
with a dot denoting differentiation with respect to $\tau$.  Since the normal
has unit length, $g^{ab}N_aN_b=1$, we can express $\dot T$ in terms of $\dot
R$,
$$\dot T = {(\dot R^2+u(r))^{1/2}\over u(r)} \neqno\dott$$ 
in the interior and with $u(r)$ replaced by $v(r)$ in the exterior bulk.  With
the normal in this form we find that the induced metric on the brane is
already in the standard Robertson-Walker form; the metric induced from the
interior bulk metric is
$$\eqalign{ ds^2  &= - u(r)\, dt^2 + (u(r))^{-1}\, dr^2 
+ R^2\, d\Omega^2_3 \cr 
&= - u(r)\left( \dot T^2 - (u(r))^{-2}\dot R^2\right) \, d\tau^2 
+ R^2\, d\Omega^2_3 \cr 
&= - d\tau^2 + R^2(\tau)\, d\Omega^2_3 \cr
&\equiv h_{\mu\nu}\, dx^\mu dx^\nu ,\cr} \neqno\induced$$ 
where $\mu$, $\nu$ run over the coordinates on the brane.  The exterior region
produces exactly the same induced metric.  

In terms of the coordinate system defined by $\induced$, the interior
contribution to the extrinsic curvature is
$$K^{(1)}_{\mu\nu}\, dx^\mu dx^\nu = -{1\over u(R(\tau))\dot T} 
\left[ \ddot R + {1\over 2}{\partial u\over \partial R} \right]\, d\tau^2 
+ u(R(\tau))\dot T R\, d\Omega^2_3  \neqno\extrfour$$ 
with the exterior region contributing an analogous expression with
$u(R(\tau))$ replaced with $v(R(\tau))$.

Let us consider the matter on the brane to be distributed as an isotropic
perfect fluid, of density $\rho$ and pressure $p$, for which the
energy-momentum tensor is
$$T_\mu^\nu = \hbox{diag}(-\rho,p,p,p). \neqno\fluidse$$ 
In this case, the spatial components of $\jumptwo$ together with $\dott$ yield 
$$\sqrt{\dot R^2+u(R)} \pm \sqrt{\dot R^2+v(R)} 
= {R\over\ell} {\sigma\over\sigma_c} - {b\ell\over 2R} 
\left[ \dot R^2 + k \right] + {8\pi G\over 3} R \rho + \cdots .
\neqno\israel$$
The temporal component of $\jumptwo$ does not give an independent equation
once we have imposed the conservation of energy on the brane $\per$ $\garriga$
which demands that
$${d\over d\tau} \left( \rho R^3 \right) = - p {d\over d\tau}
R^3. \neqno\conserve$$ 
The choice of the relative sign between the extrinsic curvature terms in
$\israel$ depends on the geometry of the bulk AdS$_5$ space that surrounds the
brane.  In the original RS universe, the orbifold is made of two slices of
AdS$_5$ attached so that the warp factor---the $r^2/\ell^2$ in the AdS metric
$\adsbulk$---decreases as we move further from the brane in either direction. 
Thus for the orbifold geometry, the plus sign is chosen.  When the warp factor
behaves differently on opposite sides of the brane, as for a brane simply
embedded in a single bulk AdS$_5$ space, the minus sign is used.

For no scalar curvature term, $b=0$, the Israel condition $\israel$ can be
rewritten so that the evolution of $R(\tau)$ is determined by a potential,
$$\half \dot R^2 + V(R) = - \half k , \neqno\potdefi$$ 
where
$$\eqalign{ V(R)  
&= - {1\over 8} {R^2\over\ell^2} \left\{ {(\sigma+\rho)^2\over\sigma_c^2}  
   - 2 \left( {\ell^2\over\ell_1^2} + {\ell^2\over\ell_2^2} \right)
   + {\sigma_c^2\over(\sigma+\rho)^2} 
   \left( {\ell^2\over\ell_1^2} - {\ell^2\over\ell_2^2} \right)^2 \right\} \cr
&\quad  - {1\over 4}
{1\over R^2} \left\{ m_1 + m_2 
   - {\sigma_c^2\over(\sigma+\rho)^2} (m_1-m_2) 
     \left( {\ell^2\over\ell_1^2} - {\ell^2\over\ell_2^2} \right) \right\} \cr
&\quad  - {1\over 8}
{\sigma_c^2\over(\sigma+\rho)^2} {\ell^2 (m_1-m_2)^2\over R^6} . \cr} 
\neqno\potential$$ 
A similar potential is implicit in $\per$.  For $R\gg\ell$ and for a generic
tension, this potential does not produce a Standard Robertson-Walker
cosmology.  However, when the brane tension is tuned to 
$$\sigma = \pm\sigma_c \left| {\ell\over\ell_1} \pm {\ell\over\ell_2} \right|
\neqno\crittension$$ 
the leading $R^2/\ell^2$, $\rho$-independent term drops out of the potential. 
The appropriate signs in $\crittension$ depend on the behavior of the AdS
space on either side of the brane.

The simplest example of a system that produces a realistic cosmology is an
AdS$_5$ space that terminates on a `edge of the universe' $3$-brane, in the
spirit of
\ref\horava{P.~Horava and E.~Witten, ``Eleven-Dimensional Supergravity on a
Manifold with Boundary,'' Nucl.\ Phys.\ {\bf B475}, 94 (1996)
[hep-th/9603142].},
with only a tension term in the brane action.  This closely resembles the
usual orbifold geometry $\per$ except that here the critical tension,
$\sigma=\sigma_c$, is half of that needed for the orbifold.  For
$\ell_1=\ell$, $\ell_2\to\infty$ and $m_1=m$, $m_2=0$, the potential
$\potential$ becomes
$$V(R) = -{1\over 2} {R^2\over\ell^2} {1\over\sigma_c^2} \left[
(\sigma+\rho)^2 - \sigma_c^2 \right] - {m\over 2R^2}. \neqno\potnok$$
Making the fine-tuning of the brane tension to its critical value,
$$\dot R^2 + k = {R^2\over\ell^2} {2\rho\over\sigma_c} +
{R^2\over\ell^2} {\rho^2\over\sigma_c^2} + {m\over R^2} = {16\pi
G\over 3\ell} \rho R^2 + {m\over R^2} + \cdots .
\neqno\potnokcrit$$ 
Here we have inserted the definition of $\sigma_c$ and assumed
$\rho/\sigma_c\ll 1$.  For the standard Robertson-Walker universe, the
dynamical equation that determines $R(\tau)$ is
$$\dot R^2 + k = {8\pi G_4\over 3}\rho R^2 \neqno\stdrw$$ 
where $G_4$ is the $3+1$ dimensional Newton's constant.  Thus identifying $G_4
= 2G/\ell$, we recover the familiar cosmologies on the brane driven by the
energy density on the brane, as long as $m$ is not too large.  A similar
result was found in this edge of the universe picture in 
\ref\gubser{S.~S.~Gubser, ``AdS/CFT and gravity,'' hep-th/9912001.}.  

\newsec{A Vacuum Bubble.}

When a bubble nucleates in a region having a vacuum energy higher than that in
the bubble's interior, the bubble will expand or contract depending upon the
surface tension of the bubble and the difference in the bulk vacuum energies. 
A simple example of this behavior occurs when a bubble of AdS$_5$ is
surrounded by an asymptotically flat region.  The $3$-brane here is the
surface of this bubble.  The purpose of this section is to introduce this
bubble as an example of an acceptable brane cosmology that is driven by one of
the mass parameters in the bulk metric in a relatively simple setting.  One
obvious difficulty---that the model does not produce a $4d$ Newton's Law---can
be removed by adding a scalar curvature term to the brane action.  Yet we
shall first study the cosmology without this term since in this limit we can
solve the behavior exactly and shall find that it is maintained when the brane
curvature term is included.

For a bubble in a flat vacuum, the metrics for the interior and exterior
regions are then respectively given by
$$u(R) = {r^2\over \ell^2} + k - {m_1\over r^2} 
   \qquad\qquad       v(R) = k - {m_2\over r^2} .
\neqno\bubblemetric$$ 
Since $\ell_2\rightarrow\infty$, we have set $\ell_1=\ell$ without loss of
generality.  Then the cosmological evolution is determined by the function,
$$\eqalign{ V(R)  &= {1\over 8} {R^2\over\ell^2} 
   \left\{ 2 - {(\sigma+\rho)^2\over\sigma_c^2} 
   - {\sigma_c^2\over(\sigma+\rho)^2} \right\} \cr 
&\quad  - {1\over 4} {1\over R^2} \left\{ m_1 + m_2 
   - {\sigma_c^2 (m_1-m_2)\over(\sigma+\rho)^2}  
      \right\}  
- {1\over 8} {\sigma_c^2\ell^2 (m_1-m_2)^2 \over(\sigma+\rho)^2 R^6} . \cr} 
\neqno\bubbleone$$ 
Again, this potential does not to lead to a standard Robertson-Walker
cosmology on the brane unless we set $\sigma=\sigma_c$.  Expanding in the
limit where the matter density is small compared to this critical tension, we
have
$$\eqalign{ V(R) &= - {1\over 2} {m_2\over R^2} - {1\over 8}
{\ell^2(m_2-m_1)^2\over R^6} + \cdots \cr &\quad + {1\over 2}
{\rho\over\sigma_c} \left( {m_2-m_1\over R^2} + {1\over 2}
{\ell^2(m_2-m_1)^2\over R^6} + \cdots \right) \cr &\quad - {1\over 2}
{\rho^2\over\sigma_c^2} \left( {R^2\over\ell^2} + {3\over 2}
{m_2-m_1\over R^2} + {3\over 4} {\ell^2(m_2-m_1)^2\over R^6} + \cdots
\right) + \cdots. \cr} \neqno\bubbletwo$$

The potential for the vacuum bubble $\bubbletwo$ does not contain a $\rho R^2$
term, since the same fine tuning that removes the cosmological constant from
the brane also eliminates such a term.  However, in the limit in which
$R(\tau)\gg\ell$ and $\rho\ll\sigma_c$, the leading term that determines the
cosmology on the surface of the expanding bubble is 
$$\dot R^2 + k = {m_2\over R^2} + \cdots .\neqno\bubblethree$$
Although this equation seems quite different from $\stdrw$, the time
dependence of its solution is exactly the same as for a radiation-dominated
universe in which $\rho\propto R^{-4}$.  Notice that if we do not want the
$\rho^2 R^2$ term to dominate, we should only consider sufficiently late times
in the evolution when
$${\rho\over\sigma_c} \ll {\ell\over R} . \neqno\mconstraint$$

\newsec{Effects of the Scalar Curvature in the Brane Action.}

\subsec{The Edge of the Universe.}

We now study the effects of including a scalar curvature term for the induced
metric in the brane action, $b\not=0$.  We first consider a bulk AdS$_5$ space
that terminates on a $3$-brane.  The Israel equation $\israel$ in this case
contains only one extrinsic curvature term,
$$- \left[ \dot R^2 + u(R) \right]^{1/2} = {R\over\ell}
{\sigma+\rho\over\sigma_c} - b{\ell\over 2R} \left[ \dot R^2 + k \right] .
\neqno\edge$$ 
Solving for $\dot R^2+k$, we obtain the potential
$$V(R) = - {R^2\over b^2\ell^2} \left[ 1 + b {\sigma+\rho\over\sigma_c} \pm 
\sqrt{ 1 + b^2 + 2b {\sigma+\rho\over\sigma_c} - b^2{m\ell^2\over R^4} } 
\right] . \neqno\potk$$ 
At the critical brane tension, for the lower sign and assuming that we can
expand the square root, we find that 
$$V(R) = - {1\over 1+b} {R^2\over\ell^2} {\rho\over\sigma_c} 
- {1\over 2(1+b)} {m\over R^2} - {1\over 2 (1+b)^3}{R^2\over\ell^2}
{\rho^2\over\sigma_c^2} + \cdots . \neqno\potkcritex$$ 
This time instead of $\potnokcrit$ we have
$$\dot R^2 + k = {1\over(1+b)}\left({16\pi G\over 3\ell} \rho R^2 
+ {m\over R^2}\right) + \cdots \neqno\potkcos$$ 
We obtain the standard Robertson-Walker evolution on the brane if we identify
the four dimensional Newton's constant with 
$$G_4 = {1\over 1+b} {2\over\ell} G . \neqno\gennewton$$ 

We obtain the same result if we calculate the $G_4$ by considering variations
about the background metric and then integrating over the extra dimension, as
described in
\ref\rsa{L.~Randall and R.~Sundrum, ``A large mass hierarchy from a small
extra dimension,'' Phys.\ Rev.\ Lett.\ {\bf 83}, 3370 (1999)
[hep-ph/9905221].}
and $\rsb$.  Since a non-zero $m$ is not needed to produce the standard
cosmology $\potkcritex$, we set it to zero while determining $G_4$.  It is
also convenient, rather than working with the coordinates of $\adsbulk$, to
define a new radial coordinate through 
$$e^{2\rho/\ell} = {r^2\over\ell^2} + 1 . \neqno\rhodef$$ 
The AdS$_5$ metric then becomes
$$ds^2 = - e^{2\rho/\ell}\, dt^2 
+ {e^{2\rho/\ell} \over e^{2\rho/\ell} - 1}\, d\rho^2 
+ \left( e^{2\rho/\ell} - 1 \right)\ell^2\, d\Omega^2_3 . 
\neqno\rhometric$$ 
In the limit, $\rho\gg\ell$, this metric reduces to the simpler form
$$ds^2 \approx e^{2\rho/\ell} \left( - dt^2 + \ell^2\, d\Omega^2_3
\right) + d\rho^2 . \neqno\rhometrica$$ 
If we replace the metric on the brane $- dt^2 + \ell^2\, d\Omega^2_3$ with a
metric $\bar g_{\mu\nu}(x^\lambda)\, dx^\mu dx^\nu$ that only depends on the
coordinates on the brane, then we find that the $5d$ scalar curvature is
related to the $4d$ scalar curvature by
$$R_5 = - {20\over\ell^2} + e^{-2\rho/\ell} \bar R_4 + \cdots
. \neqno\reduceR$$ 
Integrating over the AdS$_5$ region gives then the following term in the
effective action,
$${1\over 16\pi G_4}\int_{\cal B} d^4x\, \sqrt{-h}\, \bar R_4
\equiv {1\over 16\pi G} \int_{\cal B} d^4x\, \sqrt{-h}\, {\ell\over 2} 
\bar R_4 . \neqno\effR$$
Combining this effective brane curvature induced by the bulk zero-mode with
that included in the brane action gives
$$S_{\rm eff} = {1\over 16\pi G} \int_{\cal B} d^4x\, \sqrt{-h}\, {\ell\over
2} \left( \bar R_4 + b{\cal R} \right) + \cdots \neqno\effRwithb$$
so that the effective four dimensional Newton constant gets renormalized by
the factor $1/(1+b)$.  From the vantage of an effective field theory on the
brane, for which $b$ is determined by some unknown higher energy theory, this
result shows that we recover Newtonian gravity on the brane and at the same
time the ability to generate a standard cosmological behavior on the brane, as
long as $b>-1$.  For comparison, in the standard orbifold picture, with
$\sigma=2\sigma_c$, the effective Newton constant on the brane is
$$G_4 = {1\over 2+b} {2\over\ell} G . \neqno\rsgennewton$$

For the special choice $b=-1$ when $\sigma=\sigma_c$, our effective action on
the brane corresponds to the first two terms in the brane counterterm action
of $\bk$ and $\myers$ which regularizes the bulk AdS action.  The AdS/CFT
conjecture $\adscft$ suggests that for this action, the theory of gravity in
the AdS bulk is equivalent to a conformal field theory on the boundary,
without gravity.  Indeed we find that for physical values for the matter
density $(\rho\ge 0)$ and for a positive mass parameter in the
AdS-Schwarzschild metric $(m\ge 0)$, we do not recover a realistic
cosmological evolution on the brane.  For $b=-1$ and $\sigma=\sigma_c$, the
Israel equation $\israel$ yields a complex potential for $\dot R^2+k$,
$$V(R) = {R^2\over\ell^2} \left[ {\rho\over\sigma_c} \pm 
\sqrt{ - 2 {\rho\over\sigma_c} - {m\ell^2\over R^4} } \right] ,
\neqno\potmyers$$ 
so we no longer obtain the ordinary cosmological solutions.

\subsec{The Vacuum Bubble.}

When a brane is embedded between arbitrary bulk AdS-Schwarzschild spaces and a
scalar curvature term is included in the brane action, the Israel equation
$\israel$ is a quartic polynomial in $(\dot R^2 + k)$ which becomes tractable
only for special space-time geometries or in the $R\gg\ell$ limit.  Returning
to the case of a vacuum bubble expanding into an asymptotically flat region,
with $u(r)$ and $v(r)$ as in $\bubblemetric$, we find the following leading
behavior in the $\ell/R\ll 1$ limit:
$$\eqalign{ V(R)  &\approx - {1\over 2} {m\over R^2} 
- {1\over 8} {\ell^2\over R^2} {m^2\over R^4} (b+1)^2 
+ {1\over 2} {m\over R^2} {\rho\over\sigma_c} (b+1) + \cdots \cr
&\quad  - {1\over 2} {\rho^2\over\sigma_c^2} 
\left( {R^2\over\ell^2} + {3\over 2} {m\over R^2} (b+1)^2 + \cdots \right) 
+ \cdots \cr} \neqno\potgenkflat$$ 
Here, for simplicity, we have set $m_2=m$ and $m_1=0$.  Notice that the
presence of a brane curvature term has not generated a $\rho R^2$ term. 
Therefore we still require that the cosmology is driven by the $m/R^2$ term in
order to obtain the same time evolution as in a radiation dominated universe. 
For the new $b$-dependent terms not to overwhelm the $m/R^2$ term we must
impose $b\rho/\sigma_c\ll 1$.  Comparing with the condition already imposed by
$\mconstraint$---that the $m/R^2$ term and not the $\rho^2 R^2$ term should
drive the cosmology---we see that we can accommodate a $b\ell$ up to
cosmological scales without imposing any new constraint.

A curvature term in the brane action plays a crucial role in the vacuum bubble
scenario since it produces a $4d$ Newton's law for distances along the brane
smaller than $b\ell$ 
\ref\ch{H.~Collins and B.~Holdom, ``Linearized gravity about a brane in an
asymmetric bulk,'' hep-th/0006158.}.  
A similar result is also found in 
\ref\dgp{G.~Dvali, G.~Gabadadze and M.~Porrati, ``4D gravity on a brane in 5D
Minkowski space,'' hep-th/0005016.}  
for a brane embedded with a flat bulk on both sides.  As we just have seen,
$b\ell$ can be large without affecting the cosmology.  One unpleasant feature
of this example is that while the correct Newton's law is obtained, the
effective $4d$ Einstein equation contains a term for a scalar graviton $\ch$.  

As a more realistic variation, consider a vacuum bubble that expands into
another AdS$_5$-Schwarzschild region, rather than a flat bulk.  Unlike the
standard Randall-Sundrum picture we shall let the second AdS length, $\ell_2$,
have a large macroscopic size but which is yet much smaller than the length
associated with the brane curvature:  $\ell_1,\ell \ll \ell_2\ll b\ell$.  The
leading behavior of the cosmology $\israel$ for this universe is then governed
by 
$$\eqalign{ V(R) &= 
- {1\over 2} {1\over b} {\ell_2\over\ell_1} {m_2\over R^2} 
- {1\over 2} {1\over b} {m_1\over R^2} + \cdots 
- {\rho\over\sigma_c} \left( 
{1\over b} {R^2\over\ell^2} 
- {1\over b^2} {\ell_2\over\ell} {R^2\over\ell^2} + \cdots \right) \cr 
&\quad + {\rho^2\over\sigma_c^2} \left( 
{3\over 2} {1\over b} {\ell_1\over\ell} {R^2\over\ell^2} + \cdots
\right) + \cdots . \cr} \neqno\bubblereg$$
Unlike $\potgenkflat$, the $\rho R^2$ term is again present:
$$\dot R^2 + k = {8\pi G\over 3} {2\over b\ell}\, \rho R^2 
+ {1\over b} {\ell_2\over\ell_1} {m_2\over R^2} + \cdots .
\neqno\bubbleregcos$$
Provided $m_2$ is not too large, we recover a standard Robertson-Walker
cosmology with an effective $4d$ Newton's constant, $G_4 = 2G/b\ell$.  What
has happened for this bubble is that above the AdS lengths we expect that the
bulk space produces an effectively $4d$ theory of gravity $\rsa$.  Since we
have assumed that $\ell_1,\ell_2\ll b\ell$, when we probe distances below
$\ell_1,\ell_2$ we do not observe the extra dimensions of the bulk space since
we are in the regime in which the effect of the brane curvature term
dominates.  This argument is borne out in $\ch$ where it is shown that the
effective theory of gravity on the surface is governed by a $4d$ Einstein
equation at all scales when $\ell_1,\ell_2\ll b\ell$.

\newsec{Conclusions.}

We have found that, in general, the inclusion of a scalar curvature term in
the brane action still allows us to find the standard Robertson-Walker
cosmologies for the evolution of the brane.  This standard behavior emerges
once the size of the universe has grown large in comparison to the AdS length
of the bulk space and provided that the usual fine-tuning of the effective
cosmological constant on the brane to zero has been made.  When the AdS
lengths are small, the presence of this brane scalar curvature term simply
acts to renormalize the effective Newton constant on the brane.  In the case
an `end of the universe' brane, the brane curvature does not affect the
cosmology, except when $b=-1$.

We have explored physically intuitive brane universes in which the bulk does
not have an orbifold symmetry.  In the case of a vacuum bubble expanding into
an asymptotically flat space, we encountered an intriguing example of a system
in which the existence of the bulk is crucial for the correct cosmological
evolution since the $\rho R^2$ term that usually produces a Robertson-Walker
cosmology is absent.  Instead, the cosmology, which has the same
time-dependence as a radiation dominated universe, is driven by a mass term in
the bulk Schwarzschild metric.  A scalar curvature in the brane action plays a
more important role here since it provides the only possible source for $4d$
gravity up to a scalar graviton.  We also examined a variation in which the
bubble lies between two regions with potentially very different cosmological
constants.  For such a bubble, it is possible to recover a completely standard
Robertson-Walker cosmology without constraining the bulk AdS lengths to be
below a millimeter scale, provided that the brane curvature term is
sufficiently strong.  These examples should encourage the search for novel
extra-dimensional models in which the bulk effects are not small corrections
to the standard cosmology but rather drive its evolution.

\listrefs

\bye